\documentclass[MSNbibl,nameyear,dvips]{arxstspdf}
\usepackage{flushend}
\usepackage{stfloats}

\volume{26}
\issue{2}
\pubyear{2011}
\firstpage{266}
\lastpage{270}
\doi{10.1214/11-STS346REJ}
\referstodoi{10.1214/10-STS346}

\begin{document}
\begin{frontmatter}

\title{Rejoinder}
\runtitle{Rejoinder}

\begin{aug}
\author[a]{\fnms{J. N. K.} \snm{Rao}\corref{}\ead[label=e1]{jrao@math.carleton.ca}}
\runauthor{J. N. K. Rao}

\affiliation{Carleton University}

\address[a]{J. N. K. Rao is Distinguished Research Professor, School of
Mathematics and Statistics, Carleton University, Ottawa, Ontario K1S
5B6, Canada \printead{e1}.}

\end{aug}


\vspace*{-6pt}
\end{frontmatter}

First, I would like to thank the three discussants (Glen Meeden, Joe
Sedransk and Eric Slud) for constructive comments on my paper and for
providing additional relevant references, particularly on frequentist
model diagnostics (Slud) and Bayesian model checking (Sedransk). I
totally agree with Sedransk that studying alternative methods of making
inference for finite populations is an ``underserved field of
research.'' I will first address the constructive comments of the
discussants on the comparison of methods for handling sampling errors in
the context of estimation with fairly large domain samples.
Subsequently, I will respond to the discussions on small area
estimation.
\vspace*{-1pt}

\section*{Hansen et al. Example}
\vspace*{-1pt}

In Section 3.2, I cited the well-known Hansen,~Ma\-dow and Tepping (HMT)
example illustrating the dangers of using model-dependent methods with
fair\-ly large samples even under minor model misspecifications. Sedransk
argues in his discussion that new advances in model diagnostics, such as
model averaging, might remedy the difficulty noted by HMT and provide
improvements over the ``straw man, the usual ratio estimator.'' I agree
with Sedransk that it
would be worthwhile analyzing this example and other examples to show
how one can make valid model-dependent inferences routinely with fairly
lar\-ge domain samples that can provide significant improvements over the
design-based (possibly model-assisted) methods, particularly in the
context of official statistics with many variables of interest. If this
goal can be achieved, then I believe model-dependent methods
(frequentist or Bayesian) will have significant impact on practice,
similar to their current use in small area estimation with small domain
samples. The HMT example showed the\vadjust{\eject} importance of using design weights
under their design with deep stratification by size and disproportional
sample allocation. The usual design unbiased weighted estimator is
almost as efficient as the usual combined weighted ratio estimator under
the HMT design because of deep stratification by size, so I~do not agree
with Sedransk's comment on the importance of ratio estimator in the HMT
example. It is interesting to note that under proportional sample
allocation, the BLUP estimator (unweighted ratio estimator) under the
incorrectly specified ratio model is identical to the combined weighted
ratio estimator and hence it performs well because it is design
consistent, unlike under disproportional sample allocation. The HMT
example demonstrated the importance of design consistency, and in fact
as noted in Section 3.2, Little (\citeyear{LIT83}) proposed restricting attention to
models that hold for the sample and for which the corresponding BLUP
estimator is design consistent. I have noted some limitations of this
proposal in Section 3.2. It should be noted that the HMT illustration of
the poor performance of the BLUP estimator used the repeated sampling
design-based approach to evaluate confidence interval coverage. On the
other hand, model-based inference~is based on the distribution induced
by the model conditional on the particular sample that has been drawn.
However, Rao (\citeyear{Rao97}) showed that the HMT conclusions still hold in the
conditional framework because of the effective use of size information
through size stratification.

\section*{Role of Design Weights}

I will now turn to Meeden's useful
comments on the role of design weights and the use of Polya posterior
(PP) for making inferences after the sample is observed. As noted in
Section 4.2, the PP approach when applicable permits routine interval
estimation for any finite population parameter of interest through
simulation of many finite populations from PP and this general interval
estimation feature of PP is indeed attractive. Meeden notes in his
discussion that an R package is also available for simulating many
complete populations. However, so far the PP methodology considered only
simple designs that may satisfy the \mbox{assumption} that the un-sampled units
are like the sampled units\vadjust{\eject} (exchangeability) which limits its
applicability in practice. Meeden agrees with my comment that the PP
approach needs extension to more complex designs \mbox{before} it becomes
attractive to users. Even for the simple designs where it is applicable,
it would be useful to identify scenarios where the PP can perform
significantly better than the routine design-based methods in terms of
confidence interval coverage, especially in cases where the traditional
methods do not perform well; for example, the Woodruff interval on
quantiles under size stratification noted in Section 1. Meeden notes the
work of Lazar, Meeden and Nelson (\citeyear{l-m-n08}) on the constrained PP which incorporates
known population information about auxiliary variables without any model
assumptions about how the auxiliary variables are related to the
variables of interest, similar to calibration estimation. It appears
that the constraints allowed by this method are more flexible than those
in the usual calibration estimation, such as the population median falls
in some known interval, and this feature might prove attractive to the
user, especially due to the availability of an R package. However, the
constrained PP could run into problems when the number of population
constraints is large, similar to traditional calibration estimation.

In his concluding remarks, Meeden says that one should not focus on
estimating the variance of an estimator, but this is a customary
practice as it allows reporting estimated coefficient of variation (CV)
of the estimator as a quality measure and the user can compute
confidence interval from this variance estimator for any desired
confidence level using normal approximation. Meeden also expresses
concerns that the frequentist practice is often ``obscured by the
prominent and unnecessary role played by the design weights after the
sample has been selected.'' But design weights or calibration weights
are needed for asymptotically valid design-based inferences, although it
is often necessary to modify the weights to handle special situations,
such as outlier weights. In fact, the PP-based \mbox{estimators} of a
population mean are often close to the traditional weighted estimators,
for example under stratified random sampling.

\section*{Calibration Estimators}

Slud and I seem to agree on the limitations of model-dependent
approaches (frequentist or Baye\-sian) when the sample size in a domain of
interest is sufficiently large: possible design inconsistency of the
resulting estimators under minor model\vadjust{\eject} misspecifications, leading to
erroneous inferences. In Section~3.1 I noted the popularity of
model-free calibration estimators in the large-scale production of
official statistics from complex surveys because of their ability to
produce common calibration weights and accommodate arbitrary number of
user-specified calibration constraints. In practice, design weights are
adjusted first for unit nonresponse and then calibrated to known
user-specified totals. The calibration weights are often modified to
satisfy specified range restrictions and calibration constraints
simultaneously, but there is no guarantee that such modified weights can
be found. Rao and Singh (\citeyear{RAOSIN}, \citeyear{RAOSIN09}) proposed a ``ridge shrinkage''
approach (assuming complete response) to get around the latter problem
by relaxing some calibration constraints incrementally while satisfying
the range restrictions. Slud mentions a new method he has developed
recently (Slud and Thibaudeau, \citeyear{SluThi}) that can do simultaneous weight
adjustment for nonresponse, calibration and weight compression. This
method looks very interesting and his empirical results are encouraging.
But a solution satisfying specified range restrictions on the weights
may not exist and it would be interesting to extend the Rao--Singh
approach to handle simultaneous nonresponse adjustment and calibration.

I agree with Slud that if the weights and calibration totals are
correctly specified, the resulting calibration estimator is design
consistent even if the underlying working linear regression model uses
an incorrect or incomplete set of predictor variables, as in the example
of Section 3.1. The effect of gross misspecification of the working
model is on the coverage performance of the associated confidence
intervals and hence it is ``more subtle than design-consistency'' as
noted by Slud. Incidentally, Dorfman (\citeyear{autokey2}) used this example to
question the contention of Hansen and Tepping (\citeyear{HANTEP90}) that ``design-based
estimators that happen to incorporate a model are inferentially
satisfactory, despite failure of the model'' and concluded that the
results on coverage for the linear regression estimator calibrated on
the population size $N$ and the population total $X$ ``dramatically call
this contention into question.'' Dorfman's statement may be correct in
regard to calibration estimators based solely on user-specified
totals $Z$, but as noted in Section 3.1 a model-assisted approach based
on a working model obtained after some model checking to eliminate
gross
misspecification of the working model can lead to good confidence
interval coverage in the Dorfman example.\vadjust{\eject}

\section*{Analysis of Survey Data}

Section 3.3 of my paper on the analysis
of complex survey data is somewhat brief due to my focus on estimating
totals and means, but I should have mentioned goodness-of-fit tests that
take account of survey design. I am thankful to Slud for pointing this
out and making reference to my own work (Rao and Scott, \citeyear{RaoSco84}) on
goodness-of-fit chi-squared tests for cross-classified survey data based
on log-linear models. I might add that Roberts, Rao and Kumar (\citeyear{RobRaoKum87})
considered goodness-of-fit tests of logistic regression models with
categorical predictor variables and binary response. Graubard, Korn and
Midthune (\citeyear{GRA}) extended the well-known Hosmer and Lemeshow (\citeyear{HOSLEM80})
grouping method of goodness-of-fit for logistic regression to complex
survey data. Roberts, Ren and Rao (\citeyear{ROB}) studied goodness-of-fit tests
for mean specification in marginal models for longitudinal survey data
and obtained an adjusted Hosmer and Lemeshow test using Rao--Scott
corrections as well as a quasi-score test obtained by extending the
method of Horton et al. (\citeyear{HORetal99}) to survey data.

Multilevel models for analysis of survey data are more complex than the
marginal models for estimating regression parameters because of the
presence of random effects in the models. Goodness-of-fit methods for
two-level models, when the model holds for the sample, are available in
the literature (e.g., Pan and Lin, \citeyear{PanLin05}) but very little is known for
survey data in the presence of sample selection bias. I am presently
studying model-checking methods for two-level models taking account of
the survey design.

\section*{Small Area Estimation}

Turning now to small area estimation, Slud notes ``But one serious
objection is that each response~va\-riable would require its own Bayesian
model'' unlike direct calibration estimators using common weights. Yet
model-dependent small area methods (either HB or EB) are gaining
acceptability because direct calibration estimators are unreliable due
to small sample sizes. However, practitioners often prefer benchmarking
the small area estimators to agree with a~reliable direct calibration
estimator at a higher level.

Sedransk notes that ``almost all of the
applications use an area-level model'' even though it makes strong
assumptions such as known sampling variances, as noted in Section 5. I
agree with him that the quality of the smoothing methods used in
practice to get around the assumption of known sampling variances is
questionable although smoothed sampling variance estimates may be
satisfactory for point estimation. However, as noted in Section 5,
area-level models remain attractive because the sampling design is taken
into account through the direct estimators, and the direct estimators
and the associated area-level covariates are more readily available to
the users than the corresponding unit-level sample data. Also, in using
unit-level models one need to ensure that the population model holds for
the sample and this could be problematic, although more complex methods
have been proposed recently to handle sample selection bias in
unit-level models (Pfeffermann and Sverchkov, \citeyear{PfeSve07}). Nevertheless,
I~agree with Sedransk that unit-level models should receive more attention
in the future.

Turning to HB model diagnostics, I have noted in Section 5 some
difficulties with the commonly used posterior predictive $p$-value (PPP)
for checking goodness-of-fit of a model because of ``double use'' of
data. Alternative methods that have been proposed to avoid double use of
data are more difficult to implement, especially in the context of small
area models as noted. Sedransk mentioned three additional references
(Yan and Sedransk, \citeyear{YANSED06}, \citeyear{YanSed07}, \citeyear{YANSED10}) that studied alternative measures
in the context of detecting unknown hierarchical structures under
somewhat simplified assumptions. In particular, Yan and Sedransk
demonstrated that the unit-specific PPP-values act like uniformly
distributed random variables under the simple mean null model (without
random area effects) and hence a Q--Q plot should reveal departures from
the model. They assumed normality and absence of outliers in their
study, but it would be interesting to see if their unit-specific
P-values can in fact detect nonnormality of random effects, studied by
Sinharay and Stern (\citeyear{SinSte03}). The use of unit-specific PPP-values might be
more attractive than using the traditional PPP-function because it does
not require the selection of an appropriate checking function, but
further work is needed including the detection of nonnormality as noted
above. Yan and Sedransk showed that the PPP-function, based on the
F-statistic as the checking function, is very effective for detecting
hierarchical structure when the true model is correctly guessed as the
mean model with random area effects. This seems to imply that the
PPP-function is chosen to reject the null model and yet Sedransk
criticizes the frequentist goodness-of-fit tests by saying that\vadjust{\eject} ``such
tests are constructed to \textit{reject} null hypotheses whereas one
would like to accept a postulated model if the data are concordant with
it.'' In the simulation study of Yan and Sedransk (\citeyear{YanSed07}) the F-statistic
based PPP-value detected even small correlations when the sample size is
large and the corresponding frequentist test would also lead to similar
results. I do not agree with Sedransk that global frequentist
goodness-of-fit tests necessarily reject the null model when the data
are concordant with the model. In fact, many published papers have
identified models from real data, using frequentist tests. For example,
Datta, Hall and Mandal (\citeyear{DATHALMAN}) developed a frequentist model selection
method by testing for the presence of small area random effects and
applied the method to two real data sets involving 13 and 23 areas,
respectively. Their test is based on simple bootstrap methods and it is
free of normality assumption. The null model in both applications is a
regression model without random area effects and they showed that the
frequentist $p$-value is as large as 0.2, suggesting that the data are
concordant with the simpler null model. Slud mentioned the work of
Jiang, Lahiri and Wu (\citeyear{JiaLahWu01}) and Jiang (\citeyear{Jia01}) on mixed linear model
diagnostics in the frequentist framework. I personally prefer using
prior-free frequentist methods for model checking because they can
handle a variety of model deviations including selection of variables
and random effects selection in linear or generalized linear mixed
models (e.g., Jiang et al., \citeyear{Jiaetal08}) and detection of outliers in
multilevel models (Shi and Chen, \citeyear{ShiChe08}). A model selected by the
frequentist methods can be further subjected to Bayesian selection
methods if necessary before using HB methods for inference. Slud notes
difficulties with model checking in the context of SAIPE for sample
counties where no poor children were seen. This is also the case for
counties or areas not sampled. Model checking in those cases is indeed
challenging.

Finally, Slud makes an important observation on goodness-of-fit tests
when the primary interest is prediction: ``excellent predictions can be
provided through estimating models which are too simple to pass
goodness-of-fit checks.'' Slud notes that this observation ``has not yet
been formulated with mathematical care'' and that both frequentists and
Baye\-sians will benefit by characterizing ``which target parameters and
which combinations of true and oversimplified models could work in this
way.'' In this context, the recent work of Jiang, Nguyen and Rao (\citeyear{JIANGURAO})
on best predictive small area estimation is relevant. This paper
develops a new prediction procedure, called observed best prediction
(OBP), and shows that it can significantly outperform the traditional
EBLUP.

\section*{Acknowledgments}

Again, I am thankful to the discussants for their insightful comments. I
also wish to thank the guest editor, Partha Lahiri, for inviting me to
submit this paper to \textit{Statistical Science}. This work was
supported by a research grant from the Natural Sciences and Engineering
Research Council of Canada.

\vspace*{-4pt}

\end{document}